\documentclass[reprintnumbers,superscriptaddress,aps,pra]{revtex4}
\usepackage{dcolumn}
\usepackage{bm}
\usepackage{graphicx}
\usepackage{amsmath}
\usepackage{amssymb}
\usepackage{latexsym}
\usepackage{amsfonts}
\usepackage{array}
\usepackage{epsfig}
\usepackage{txfonts}
\usepackage{color}
\usepackage[colorlinks=true,linkcolor=blue,urlcolor=blue,citecolor=blue]{hyperref}
\usepackage{setspace}
\usepackage{braket}
\usepackage[utf8]{inputenc}
\usepackage[percent]{overpic}
\usepackage{xcolor}
\usepackage{lineno}

\newcommand{\beq}{\begin{equation}}
\newcommand{\eeq}{\end{equation}}
\newcommand{\beqa}{\begin{eqnarray}}
\newcommand{\eeqa}{\end{eqnarray}}

\begin{document}

\title{Coupled density-spin Bose-Einstein condensates dynamics and collapse in systems
with quintic nonlinearity}
\author{Jing Li}
\affiliation{Department of Physics, Shanghai University, 200444 Shanghai,
People's Republic of China \\ and International Center of Quantum Artificial Intelligence for Science and Technology (QuArtist) }
\affiliation{Okinawa Institute of Science and
Technology Graduate University, Okinawa, 904-0495, Japan}
\author{Boris A. Malomed}
\affiliation{Department of Physical Electronics, School of Electrical Engineering, Faculty of Engineering,
Tel Aviv University, Tel Aviv 6997801, Israel; Laboratory of Nonlinear-Optical Informatics,
ITMO University, St. Petersburg 197101, Russia}
\author{Wenliang Li}
\affiliation{Okinawa Institute of Science and Technology Graduate
University, Okinawa, 904-0495, Japan}
\author{Xi Chen}
\affiliation{Department of Physics, Shanghai University, 200444 Shanghai,
People's Republic of China \\ and International Center of Quantum Artificial Intelligence for Science and Technology (QuArtist) }
\affiliation{Department of Physical Chemistry, University of the Basque
Country, 48080 Bilbao, Spain}

\author{E. Ya Sherman}
\affiliation{Department of Physical Chemistry, University of the Basque
Country, 48080 Bilbao, Spain}
\affiliation{IKERBASQUE Basque Foundation for Science, Bilbao, Spain}

\date{\today }

\begin{abstract}
We investigate the effects of spin-orbit coupling and Zeeman splitting
on the coupled density-spin dynamics and collapse of the Bose-Einstein condensate
driven by the quintic self-attraction in the same- and cross-spin channels.
The characteristic feature of the collapse is the
decrease in the width as given by the participation ratio of the density rather than by the expectation
values of the coordinate. Qualitative arguments and numerical simulations reveal the existence of a
critical spin-orbit coupling strength which either prohibits or leads to the collapse, and its dependence on
other parameters, such as the condensate's norm, spin-dependent nonlinear coupling, and the Zeeman splitting.
The entire nonlinear dynamics critically depends on the initial spin sate.
\end{abstract}


\maketitle

\section{Introduction}

The idea of the collapse as a trend to catastrophic shrinkage of a
self-attracting system has proved its relevance in many branches of
nonlinear physics \cite{Sulem1999,Fibich}.
The realization of the ultracold atomic matter has greatly increased the variety of possible
nonlinear phenomena \cite{Carr2009}. In atomic Bose-Einstein condensates (BECs), 
the self-attraction driving collapse either occurs naturally, or
can be achieved by means of the engineered Feshbach resonance \cite{Cornish2000}. 
Recently developed techniques which
make it possible to produce synthetic gauge fields \cite{spielman2009,
Dalibard2011} and spin-orbit coupling (SOC) \cite{wang2010, spielman2011,
Zhai2012,Spielman2013,Zhang2016} greatly expands the versatility of the
self-interacting quantum matter and variety of collapse-related phenomena
\cite{Konotop2005,Dias2016}, as well as general properties of nonlinear two-component systems \cite{WangKdv}.
By introducing coupled spin-mass-density
dynamics with a spin-dependent velocity, SOC can affect the collapse \cite%
{Mardonov2015,Yu2017} and produce soliton-like stable states \cite%
{Sakaguchi2014,Sakaguchi2016}, which would be unstable, or would not exists,
in the absence of SOC.

In effectively one-dimensional (1D) settings, the collapse is driven by
three-body attractive interactions, which correspond to the quintic
nonlinearity in the respective Gross-Pitaevskii equations (GPEs), as the 1D
collapse cannot be caused by two-body attraction (represented by cubic terms
in the GPEs) \cite{Konler2002}. Indeed, an elementary estimate demonstrates
that the absolute values of the negative three-body energy, which drives the
self-compression of the condensate, exceeds its kinetic energy, which
impedes the collapse, by a factor $\sim N^{2},$ where $N$ is the
condensate's norm. Thus, the collapse may set in if $N$ is large enough.

Here we consider effects of SOC and Zeeman splitting (ZS) on the collapse in
a 1D system with various forms of the quintic self-attraction \cite%
{Xi2016,Astrakharchik2005,Chiquillo2017}. In 1D settings these effects can
be presented in simple and transparent form as a competition between the
velocity caused by self-attraction, which is generated in the collapse
process, and an anomalous SOC-induced spin-dependent velocity.

This paper is organized as follows. In Sec. \ref{oneb}, the model of the
spin-orbit and Zeeman-coupled BEC with three-body interactions is introduced. Then, in
Sec. \ref{sec:no-soc} we address the dynamics of the system with different initial
spin states in the absence of spin-related effects, by means of the
variational approximation. Section \ref{Sec:with-soc} deals with combined
SOC-ZS effects, by considering several different realizations of the model
demonstrating qualitatively different behaviors. The paper is concluded by
Sec. \ref{end}.

\section{The model and self-interaction}

\label{oneb}

We consider a quasi-1D BEC with pseudospin 1/2, subject to the action of the
artificial SOC and ZS and extended along the $x$ direction. The
corresponding two-component wave function is%
\begin{equation}
{\bm\psi }\left( x,t\right) =\left(
\begin{array}{c}
u(x,t) \\
v(x,t)%
\end{array}%
\right) \,,
\end{equation}%
{with total norm}
\begin{equation}
N\equiv \int_{-\infty }^{+\infty }\left( |u|^{2}+|v|^{2}\right) dx\,.
\end{equation}%
For brevity, the explicit $(x,t)$ dependence is written only when it is
necessary. The combination of same- and cross-spin interaction energies is
defined as (cf. Refs. \cite{Maim,Abdullaev2005})
\begin{equation}
E_{g}=-\frac{1}{3}\int_{-\infty }^{+\infty }\left[ g_{1}\left( \left\vert
u\right\vert ^{6}+\left\vert v\right\vert ^{6}\right) +3g_{2}\left\vert
u\right\vert ^{2}\left\vert v\right\vert ^{2}\left( \left\vert u\right\vert
^{2}+\left\vert v\right\vert ^{2}\right) \right] dx\,,  \label{Eg}
\end{equation}%
where positive coupling constants $g_{1}$ and $g_{2}$ represent the
three-particle attraction in the same- and cross-spin channels, respectively.

The evolution of the system is governed by the equation
\begin{equation}
i\hbar \frac{\partial {\bm\psi }}{\partial t}=H{\bm\psi }\,,  \label{evolution}
\end{equation}
with Hamiltonian%
\begin{equation}
H=\frac{p^{2}}{2M}+H_{\mathrm{so}}+H_{Z}+H_{g}\,,  \label{eq:Hspin}
\end{equation}%
where $p=-i\hbar \partial /\partial x$ is the momentum and $M$ the atomic
mass, while the SOC and ZS terms are%
\begin{equation}
H_{\mathrm{so}}=\frac{\alpha }{\hbar }p\,\sigma _{x},\qquad H_{Z}=\frac{\Delta }{2}%
\sigma _{z}\,.  \label{eq:HsoHZ}
\end{equation}%
Here $\alpha $ and $\Delta $ are the SOC and ZS strengths, respectively, with $%
\sigma _{x}$ and $\sigma_{z}$ being the Pauli matrices.
The synthetic spin-orbit interaction in cold atoms can be generated by Raman
coupling schemes which simultaneously flip atomic pseudospin and transfer momentum  \cite{spielman2009}.
As a result, a variety of synthetic fields can be engineered using highly coherent laser beams.
The synthetic Zeeman fields can be produced with the same experimental setup \cite{Dalibard2011,Zhai2012,Spielman2013}. 
The self-interaction term has the form
\begin{equation}
H_{g}=\left(
\begin{matrix}
{\delta E_{g}}/{\delta u^{\ast }} & 0 \\
0 & {\delta E_{g}}/{\delta v^{\ast }}%
\end{matrix}%
\right) ,  \label{Hg}
\end{equation}%
with variations
\begin{eqnarray}
&&\frac{\delta E_{g}}{\delta u^{\ast }}=-\left[ g_{1}\left\vert u\right\vert
^{4}+g_{2}\left( 2\left\vert uv\right\vert ^{2}+\left\vert v\right\vert
^{4}\right) \right] ,  \notag \\
&&  \label{d/d} \\
&&\frac{\delta E_{g}}{\delta v^{\ast }}=-\left[ g_{1}\left\vert v\right\vert
^{4}+g_{2}\left( 2\left\vert uv\right\vert ^{2}+\left\vert u\right\vert
^{4}\right) \right] .  \notag
\end{eqnarray}

The total energy of the system is
\begin{equation}
E=E_{k}+E_{\mathrm{so}}+E_{Z}+E_{g}\,,
\end{equation}%
where
\begin{equation}
E_{k}=\frac{1}{2M}\int_{-\infty }^{+\infty }\left( \left\vert \frac{\partial u}{%
\partial x}\right\vert ^{2}+\left\vert \frac{\partial v}{\partial x}%
\right\vert ^{2}\right) dx  \label{Ek}
\end{equation}%
is the kinetic energy, with the SOC and ZS
terms being
\begin{eqnarray}
E_{\mathrm{so}} &=&-i\alpha \int_{-\infty }^{+\infty }\,{\bm\psi }^{\dag
}\,\sigma _{x}\,\frac{\partial {\bm\psi }}{\partial x}dx\,,  \label{SO} \\
E_{Z} &=&\frac{\Delta }{2}\langle \sigma _{z}\rangle \,N\,,
\label{eq:Zeeman1}
\end{eqnarray}%
and
\begin{equation}
\langle \sigma _{z}\rangle \equiv \frac{1}{N}\int_{-\infty }^{+\infty }\,{\bm%
\psi }^{\dag }\sigma _{z}\,{\bm\psi }dx\,.  \label{sigma}
\end{equation}

By means of obvious rescaling, we set $\hbar =M=1$ and thus cast Eqs. (\ref%
{evolution})-(\ref{d/d}) into a coupled GPE system:
\begin{eqnarray}
i\frac{\partial }{\partial t}u &=&\left( -\frac{1}{2}\frac{\partial ^{2}}{%
\partial x^{2}}+\frac{\Delta }{2}+\frac{\delta E_{g}}{\delta u^{\ast }}%
\right) u-i\alpha \frac{\partial }{\partial x}v\,,  \label{eq:eq1} \\
i\frac{\partial }{\partial t}v &=&\left( -\frac{1}{2}\frac{\partial ^{2}}{%
\partial x^{2}}-\frac{\Delta }{2}+\frac{\delta E_{g}}{\delta v^{\ast }}%
\right) v-i\alpha \frac{\partial }{\partial x}u\,.  \label{eq:eq2}
\end{eqnarray}%
Note that the SOC\ Hamiltonian introduces a spin-dependent velocity defined
by commutator
\begin{equation}
V=i\left[ \frac{p^{2}}{2}+H_{\mathrm{so}},x\right] =p+V_{\mathrm{so}}\,,
\label{eq:vso}
\end{equation}%
where $V_{\mathrm{so}}=\alpha\, \sigma _{x}$ becomes a non-diagonal $2\times 2$ matrix.
As shown below, in the course of the BEC evolution, this anomalous
velocity makes its density spatially split in two spin-projected species. Depending on the
$E_{g}$ energy in Eq. \eqref{Eg}, this splitting can either prevent the
collapse, suppressing the same-spin self-interaction due to the $g_{1}$ term,
or drive the collapse by enhancing cross-spin coupling caused by the $g_{2}-$related contribution.

It is also relevant to mention that, in the absence of ZS ($\Delta =0$),
Eqs. (\ref{eq:eq1}) and (\ref{eq:eq2}) admit a reduction to the single GPE
with the quintic self-attractive term and no SOC. Indeed, the substitution
of
\begin{equation}
v\left( x,t\right) =\pm u\left( x,t\right) =U\left( x,t\right) \exp \left(
\mp i\alpha x+i\alpha ^{2}t/2\right)   \label{+-}
\end{equation}%
transforms both equations into one:%
\begin{equation}
i\frac{\partial }{\partial t}U=\left( -\frac{1}{2}\frac{\partial ^{2}}{%
\partial x^{2}}-g|U|^{4}\right) U,~g\equiv g_{1}+3g_{2}.  \label{U}
\end{equation}%
It is commonly known that Eq. (\ref{U}) admits a family of \textit{%
Townes-soliton} solutions \cite{Abdullaev2005} with arbitrary chemical
potential $\mu <0$:%
\begin{equation}
U=e^{-i\mu t}\left( -\frac{3\mu }{g}\right) ^{1/4}\sqrt{\mathrm{sech}\left( 2%
\sqrt{-2\mu }x\right) }\,.  \label{Townes}
\end{equation}%
The solution family is degenerate, as the soliton's norm does not depend on $%
\mu $:%
\begin{equation}
N_{\mathrm{Townes}}=\frac{\pi }{2}\sqrt{\frac{3}{2g}}\,,  \label{NT}
\end{equation}%
and all the solitons are unstable against the onset of the collapse.
However, the possibility of the reduction, based on Eq. (\ref{+-}), to the
single GPE (\ref{U}) does not mean that the dynamics governed by Eqs. (\ref%
{eq:eq1}) and (\ref{eq:eq2}) with $\Delta =0$ reduces to that of the
single-GPE in the generic situation. In particular, we will show that the SOC suppresses
the onset of the collapse in solutions stemming from the single-component input, given by Eq. (\ref%
{ux0vx0}), if the spin-orbit coupling $\alpha$ is larger than a critical value $\alpha_{\mathrm{cr}},$ dependent
on the system parameters $N,g_{1},g_{2},\Delta,$ and the initial conditions.

\section{Dynamics without spin-orbit coupling: the role of the spin
state}

\label{sec:no-soc}
As a starting point, we present here results for the collapse in the absence
of SOC, taking the input as an eigenstate of $\sigma _{z}$,
\begin{equation}
u(x,0)=A(0)\exp \left( -\frac{x^{2}}{2a^{2}(0)}\right) ,\quad v(x,0)=0,
\label{ux0vx0}
\end{equation}%
where $A(0)$ and $a(0)$ are the initial amplitude and width of the wave
packet, with norm
\begin{equation}
N=\sqrt{\pi }\,A^{2}(0)\,a(0)\,.  \label{eq:norm-z}
\end{equation}%
The corresponding interaction and kinetic energies, (\ref{Eg}) and (\ref{Ek}%
),\ are%
\begin{equation}
E_{g}=-\frac{1}{3}\int_{-\infty }^{+\infty }g_{1}|u(x,0)|^{6}dx=-\frac{%
g_{1}\,N^{3}}{3\sqrt{3}\,\pi \,a^{2}(0)}\,,
\end{equation}%
\begin{equation}
E_{k}=\frac{1}{2}\int_{-\infty }^{+\infty }\left\vert \frac{\partial u(x,0)}{%
\partial x}\right\vert ^{2}dx=\frac{N}{4a^{2}\left( 0\right) }\,,
\label{eq:Ekin}
\end{equation}%
whose ratio $|E_{g}/E_{k}|$ \ scales $\sim N^{2}$, as mentioned above,
implying the domination of the the interaction energy for large norms. To
study the evolution of the first component of ${\bm\psi }(x,t)$, we use the
variational ansatz,
\begin{equation}
{\bm\psi }(x,t)=A(t)\exp \left[ -\frac{x^{2}}{2a^{2}(t)}+ib(t)x^{2}\right]
\left(
\begin{array}{c}
1 \\
0%
\end{array}%
\right) \,,
\end{equation}%
where $a(t)$ and $b(t)$ are the time-dependent width and chirp,
respectively. The corresponding Euler-Lagrange equations, corresponding to
Eqs. (\ref{eq:eq1}) and (\ref{eq:eq2}), are \cite%
{Anderson,Progress}%
\begin{equation}
\frac{da}{dt}=2ab,\qquad\frac{db}{dt}=\frac{1}{2a^{4}}-2b^{2}-\frac{2g_{1}N^{2}}{3%
\sqrt{3}\,\pi a^{4}}\,.  \label{ab}
\end{equation}%
Further, after the elimination of $b$ from Eq. (\ref{ab}), one arrives at the
Ermakov equation \cite{Ermakov,Ermakov2},%
\begin{equation}
\ddot{a}=-\frac{\Lambda }{a^{3}},\qquad \Lambda \equiv \frac{4g_{1}N^{2}}{3\sqrt{3}%
\,\pi }-1\,,  \label{eq:Gaussian}
\end{equation}%
which admits well-known analytical solutions with two integration
constants, $C_{1,2}$:
\begin{equation}
a(t)=\pm \sqrt{-\Lambda /C_{1}+C_{1}\,t^{2}+2C_{1}C_{2}\,t+C_{1}C_{2}^{2}}\,.
\label{a}
\end{equation}%
For the initial condition $(da/dt)(0)=0$, the positive branch in Eq. (\ref{a})
yields
\begin{equation}
a(t)=\sqrt{-\Lambda /C_{1}+C_{1}\,t^{2}}\,.  \label{a2}
\end{equation}%
The critical norm for the Gaussian ansatz (\ref{ux0vx0}) with initial $%
\langle \sigma _{z}\rangle =1$ [see Eq. (\ref{sigma})], determined by
setting $\Lambda =0$ in Eq. (\ref{eq:Gaussian}), is
\begin{equation}
N_{c}^{\langle \sigma _{z}\rangle =1}=\sqrt{3\sqrt{3}\pi /4}\,%
g_{1}^{-1/2}\approx 2.02\,g_{1}^{-1/2}.
\label{crit}
\end{equation}%
This value is only slightly larger than the exact value of the norm of the
corresponding Townes soliton, $(\pi /2)\sqrt{3/2}\,g_{1}^{-1/2}\approx
1.924\,g_{1}^{-1/2}$, see Eq. (\ref{NT}), demonstrating that the
Gaussian ansatz is suitable for predicting the corresponding critical norm.

For $\Lambda >0$, solution (\ref{a2}) with $C_{1}<0$ describes the collapse
process \cite{Abdullaev2003}:
\begin{equation}
a(t)=a(0)\sqrt{1-\frac{t^{2}}{t_{c}^{2}}}\,,  \label{eq:at}
\end{equation}%
where $a^{2}(0)\equiv \Lambda /\left\vert C_{1}\right\vert $, and the
collapse time is $t_{c}=\sqrt{\Lambda }/\left\vert C_{1}\right\vert \equiv
a^{2}(0)/\sqrt{\Lambda }.$ The characteristic velocity developed in the
course of collapsing is
\begin{equation}
V_{c}=a(0)/t_{c}=\sqrt{\Lambda }/a(0).  \label{Vc}
\end{equation}

To estimate an effect of {the interplay of two spin components on the onset
of the collapse}, we now take {the eigenstate of $\sigma _{x}$ as the
initial state} [cf. Eq. (\ref{ux0vx0})]:%
\begin{equation}
u(x,0)=v(x,0)=\frac{A(0)}{\sqrt{2}}\exp \left( -\frac{x^{2}}{2a^{2}(0)}\right)
\label{ux0vx01}
\end{equation}%
with the same norm as in Eq. \eqref{eq:norm-z}. In this case, the kinetic
energy is again given by Eq. \eqref{eq:Ekin}, while the interaction energy
is
\begin{equation}
E_{g}=-\frac{1}{4}\,(g_{1}+3g_{2})\,\frac{\,N^{3}}{3\sqrt{3}\,\pi \,a^{2}(0)}%
\,.
\label{eq:g1g2}
\end{equation}%
Since at $g_{2}=0,$ the self-interaction energy for the state in Eq. \eqref{ux0vx01} is factor
of 2 smaller than that in Eq. \eqref{ux0vx0} while their kinetic energies are
equal, the critical value $N_{c}^{\langle \sigma
_{x}\rangle =1}$ is larger by a factor of $2$ than one given by Eq. (\ref%
{crit}). For $g_{2}=g_{1}$
(the Manakov's spin-isotropic form of the interaction \cite{Manakov1974}),
the critical norm is independent of the spin orientation.

\section{Collapse in the presence of the spin-orbit coupling and Zeeman
couplings}

\label{Sec:with-soc}

Here we focus on how the SOC and ZS strengths, $\alpha $ and $\Delta $,
respectively, affect the collapse dynamics. To this end, we perform a
qualitative analysis and numerically solve the GPE system of Eqs. (\ref%
{eq:eq1}) and (\ref{eq:eq2}), using the split-operator technique \cite%
{Chaves2015} for SOC systems.

To characterize the evolution of the width of the wave packet, we use the
time-dependent inverse participation ratio (IPR) \cite{IPR2000}, defined as
\begin{equation}
\zeta (t)\equiv \int_{-\infty }^{+\infty }\left( |u|^{2}+|v|^{2}\right)
^{2}dx.  \label{zeta}
\end{equation}%
In turn, the IPR-related width is defined as
\begin{equation}
a_{\zeta }(t)=N^{2}/\sqrt{2\pi }\,\zeta .  \label{azeta}
\end{equation}%
Another definition of the width is determined by the total spread of the
wave packet,
\begin{equation}
a_{s}^{2}(t)\equiv \langle x^{2}(t)\rangle \equiv \frac{1}{N}\int_{-\infty }^{+\infty
}x^{2}\left( |u|^{2}+|v|^{2}\right) dx.  \label{x2}
\end{equation}%
It is shown below that these widths feature qualitatively different time
dependences, due to a non-Gaussian actual shape of the collapsing wave
packet.

\subsection{Effect of the spin-orbit coupling at short time}

Here we take an eigenstate of $\sigma _{z}$ as the initial state, i.e. ${\bm%
\psi }(x,0)=[u_{0}(x),0]^{\mathrm{T}}$ ($\mathrm{T}$ stands for
transposition), and begin with an analytical consideration for small $t$,
which provides a good insight into the collapse dynamics. To single out the
qualitative SOC\ effect, we, for the time being, omit the nonlinearity (and
ZS) in Eqs. (\ref{eq:eq1}) and (\ref{eq:eq2}), thus observing that SOC
splits the input into eigenstates of $\sigma _{x}$, and these spin-projected
components evolve independently:
\begin{equation}
{\bm\psi (x,t)}=\frac{1}{2}\left(
\begin{array}{c}
u_{0}\left( x-\alpha t\right) \\
u_{0}\left( x-\alpha t\right)%
\end{array}%
\right) +\frac{1}{2}\left(
\begin{array}{c}
u_{0}\left( x+\alpha t\right) \\
-u_{0}\left( x+\alpha t\right)%
\end{array}%
\right) \,.
\label{eq:splitting}
\end{equation}

At small $t,$ using expansion
\begin{equation}
u_{0}(x\pm \alpha t)\approx u_{0}(x)\pm \alpha tu_{0}^{\prime }(x)+\frac{1}{2%
}\left( \alpha t\right) ^{2}u_{0}^{\prime \prime }(x),  \label{prime}
\end{equation}%
we identify the leading terms,
\begin{equation}
u(x,t)=u_{0}(x)+\frac{1}{2}u_{0}^{\prime \prime }(x)\left( \alpha
t\right) ^{2},
\quad
v(x,t)=-u_{0}^{\prime }(x)\alpha t.  \label{prime-prime}
\end{equation}%
The evolution produces a spin dipole moment, defined as
\begin{equation}
d_{\sigma }\equiv \frac{1}{N}\int_{-\infty }^{+\infty }\left\langle {\bm\psi
}|x\,\sigma _{x}|{\bm\psi }\,\right\rangle dx\,,
\end{equation}%
which represents separation of the spin components, and may be considered as
an additional dynamical variable. For the spinor wave function taken as per
Eqs. (\ref{eq:splitting})-(\ref{prime-prime}), one has
\begin{equation}
d_{\sigma }=\alpha \,t,
\end{equation}%
demonstrating the separation of spin contributions with opposite $\langle
\sigma _{x}\rangle $. Accordingly, the width of the wave packet, defined as
per Eq. (\ref{zeta}), is
\begin{equation}
a_{\zeta }(t)=a_{\zeta }(0)\left[ 1+4\frac{d_{\sigma }^{2}}{\zeta (0)}%
\int_{-\infty }^{+\infty }\left( u_{0}(x)u_{0}^{\prime }(x)\right) ^{2}dx%
\right] .
\end{equation}

Now one can calculate the evolution of the $z$ component of the spin,
\begin{equation}
\langle \sigma _{z}\rangle =1-4d_{\sigma }^{2}\frac{E_{k}}{N}\,.
\label{eq:Zeeman2}
\end{equation}%
If the ZS term is restored, the respective Zeeman energy shift is
\begin{equation}
E_{Z}=-2\,\Delta\, d_{\sigma }^{2}\,E_{k}.  \label{Zeeman}
\end{equation}
It depends on the sign of the ZS strength, $\Delta $, demonstrating the
significance of the direction of the Zeeman field. Further, the correction
to interaction energy (\ref{Eg}) is
\begin{equation}
\Delta E_{g}=\left( 5g_{1}-g_{2}\right) d_{\sigma }^{2}\int_{-\infty
}^{\infty }\left[ u_{0}^{\prime }(x)\right] ^{2}u_{0}^{4}(x)dx,
\end{equation}%
demonstrating that the effect of self-interaction nearly cancels at $%
5g_{1}=g_{2}$.

As an example, we take the initial state in the form of Eq. (\ref{ux0vx0}),
where $a_{\zeta }(0)=a(0)$. For the interaction-energy correction we obtain
\begin{equation}
\Delta E_{g}=\left( 5g_{1}-g_{2}\right) d_{\sigma }^{2}\frac{\sqrt{3}}{18\pi
}\frac{N^{3}}{a^{4}(0)},
\label{eq:DeltaEg1g2}
\end{equation}%
and the spin projection becomes%
\begin{equation}
\langle \sigma _{z}\rangle =1-\frac{d_{\sigma }^{2}}{a^{2}(0)}.
\end{equation}%
Accordingly, the IPR-determined packet's width increases as
\begin{equation}
a_{\zeta }(t)=a(0)\left[ 1+\frac{d_{\sigma }^{2}}{a^{2}(0)}\right] .
\end{equation}%
Comparing this expression to Eq. \eqref{eq:at}, we see that the effect of
SOC on the BEC width dominates for
\begin{equation}
\alpha >V_{c},  \label{>}
\end{equation}
i.e., if the anomalous velocity exceeds the characteristic velocity
developed in the course of collapsing.

\begin{figure}[h]
\centering
\includegraphics*[width=0.45\textwidth]{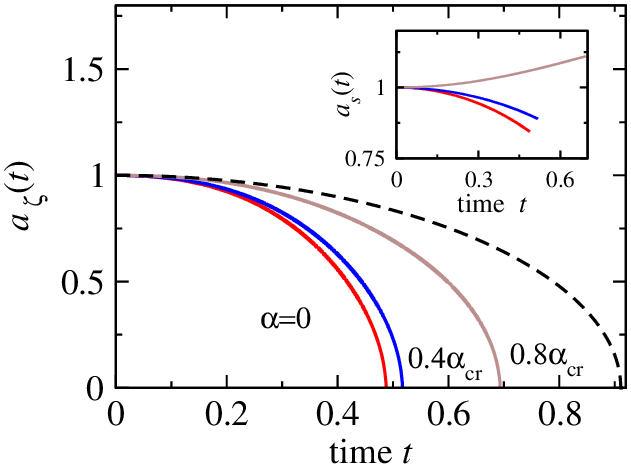}
\caption{(Color online) Widths $a_{\zeta }(t)$, calculated as per
Eq. (\ref{zeta}) (the main plot), and $a_{s}(t)$, defined as per Eq.
(\ref{x2}) (the inset), for different values of the SOC strength, $%
\alpha $, shown near the plots. Here $\alpha _{\mathrm{cr}%
}=1.16$ for $N=3$, $g_{1}=1$, $g_{2}=0$, and $\Delta =0$. For comparison,
the dashed line displays the width evolution for the Gaussian-ansatz
solution, given by Eq. \eqref{eq:at}. Note that for this choice of
parameters $\Lambda \approx 1.2$ [see Eq. (\ref{eq:Gaussian})], and
the ansatz predicts the collapse at $t_{c}\approx 0.9$ and $V_{c}\approx 1.1$%
, very close to the numerically obtained value, $\alpha _{\mathrm{cr}%
}=1.16$, as seen here. Here and in the following figures we use $a(0)=1$ for the initial Gaussian
states.}
\label{fig:collapse_1}
\end{figure}

\subsection{Zero cross-interaction: $g_{1}\neq 0,g_{2}=0,\Delta =0$}

We begin systematic analysis of the dynamics, neglecting the nonlinear
interaction between the components, i.e., setting $g_{2}=0$ in Eqs. (\ref%
{eq:eq1}) and (\ref{eq:eq2}). In Fig. \ref{fig:collapse_1}, we display
numerical results for three different values of the SOC strength, $\alpha $.
Note that the packet's shape is strongly non-Gaussian even in the case of $%
\alpha =0$. As a result, width $a_{s}(t)$, defined by Eq. (\ref{x2}), does
not display the collapse, and one needs to examine the IPR as in Eqs. \eqref{zeta} and \eqref{azeta}.
The reason for this peculiarity is discussed below.

Naturally, the collapse time strongly depends on $\alpha $, diverging when
the SOC strength is approaching a critical value, $\alpha _{\mathrm{cr}}.$
This value is determined by condition (\ref{>}), which implies that a
typical anomalous velocity induced by SOC, $V_{\mathrm{so}}\sim \alpha $,
exceed the collapse velocity $V_{c}$, see Eq. (\ref{Vc}). If $\alpha =\alpha
_{\mathrm{cr}}$, the simulations demonstrate that the BEC starts its
evolution by compressing in the beginning, but then slightly expands (not
shown in detail). As $\alpha$ approaches $\alpha_{\mathrm{cr}},$ the
collapse time diverges as $\sim (\alpha _{\mathrm{cr}}-\alpha )^{-1/2}$.


\begin{figure}[t]
\centering
\includegraphics*[width=0.45\textwidth]{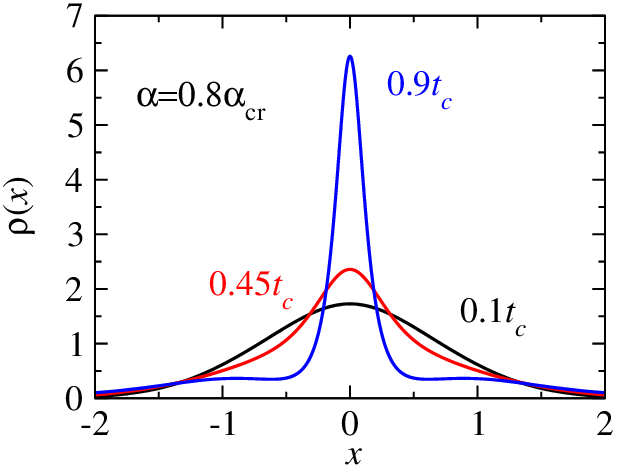}
\caption{(Color online) Density profiles of the collapsing wave packet, $%
\rho (x)=\left\vert v\right\vert ^{2}+\left\vert u\right\vert ^{2}$,
produced by numerical simulations of Eqs. (\ref{eq:eq1}) and (%
\ref{eq:eq2}) with initial conditions (\ref{ux0vx0}), at
time $t$ marked near the plots. Wings are clearly seen in the figure at $\left\vert
x\right\vert>1/2.$ Here $\alpha =0.8\alpha _{\mathrm{cr}}$
for $N=3$, $g_{1}=1$, $g_{2}=0$, and $\Delta =0$.}
\label{fig:density}
\end{figure}

To better understand the difference between the evolution of the IPR- and
the $\sqrt{\langle x^{2}\rangle }$-based widths, which are defined by Eqs. (\ref{eq:at})
and  (\ref{azeta}), we display, in Fig. \ref{fig:density},
density profiles at different times, which show well-developed wings.
Therefore, while the core part of the wave packet collapses, its shell
extends and leads to finite $\langle x^{2}\rangle .$ The evolution of the
wings, initially observed in Ref. \cite{Abdullaev2005}, is attributed to
very fast decrease in the interaction-energy density.
This is confirmed by comparing the numerical and variational solutions.
The collapse occurs faster than predicted by Eq. \eqref{eq:at}, as only a
fraction of the BEC density in the vicinity of the origin undergoes the
collapse. Thus, although the Gaussian ansatz accurately predicts the
critical norm, it is not suitable for modeling dynamical features of the
collapse.

\begin{figure}[t]
\centering
\includegraphics*[width=0.45\textwidth]{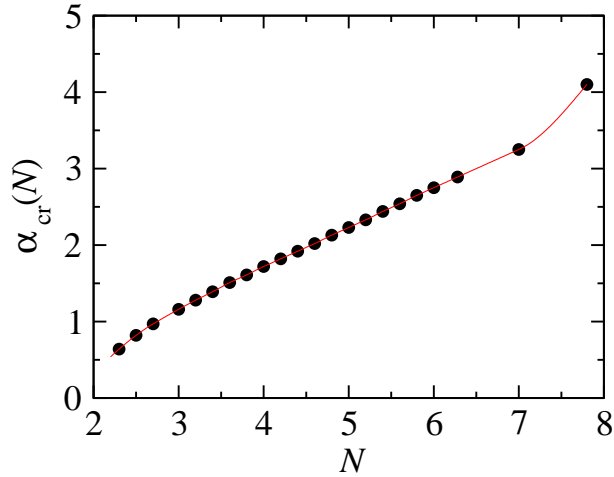}
\caption{(Color online) Critical value $\alpha _{\mathrm{cr}}$ of
the SOC strength, such that the collapse in the system with the
self-attraction ($g_{1}=1$, $g_{2}=0,\Delta =0$) does not occur at $%
\alpha >\alpha _{\mathrm{cr}}$, vs. the total norm, $N$. The red
line is an interpolation connecting all data points. Note that, as clearly
seen in the figure, the condition $\alpha _{\mathrm{cr}}\sim %
\sqrt{\Lambda }/a(0)$ predicts linear dependence of $\alpha _{%
\mathrm{cr}}(N)$ in the interval of $N_{\mathrm{cr}}^{\left\langle %
\sigma _{z}\right\rangle =1}\alt N\alt4N_{\mathrm{cr}}^{\left\langle %
\sigma _{z}\right\rangle =1}$, where $N_{c}^{\left\langle \sigma %
_{z}\right\rangle =1}\approx 2.0$ is the critical value either in the variational approximation (\ref{crit}) or
the exact one in Eq. \eqref{NT}.}
\label{fig:alphaN}
\end{figure}

As follows from condition $\alpha _{\mathrm{cr}}\sim V_{c}$, the critical
value of the SOC strongly depends on the norm of the
wave packet. For sufficiently large $N$, no value of $\alpha $ is sufficient
to arrest the collapse, as shown in Fig. \ref{fig:alphaN}, since a fully
separated packet still has a sufficient norm to drive the collapse. Taking
into account the fact that $N_{c}^{\left\langle \sigma _{x}\right\rangle
=1}=2N_{c}^{\left\langle \sigma _{z}\right\rangle =1}$, and that the fast
separation reduces the norm per component by a factor of two, we conclude
that, at $N>4N_{c}^{\left\langle \sigma _{z}\right\rangle =1}$, the critical
value $\alpha _{\mathrm{cr}}$ diverges, as can be seen in the more rapid increase in $%
\alpha _{\mathrm{cr}}$ in Fig. \ref{fig:alphaN} as $N$ approaches
$4N_{c}^{\left\langle \sigma _{z}\right\rangle =1}.$

\subsection{Effect of the Zeeman splitting: $g_{1}\neq 0,g_{2}=0,\Delta \neq
0$}

\begin{figure}[t]
\centering
\includegraphics*[width=0.45\textwidth]{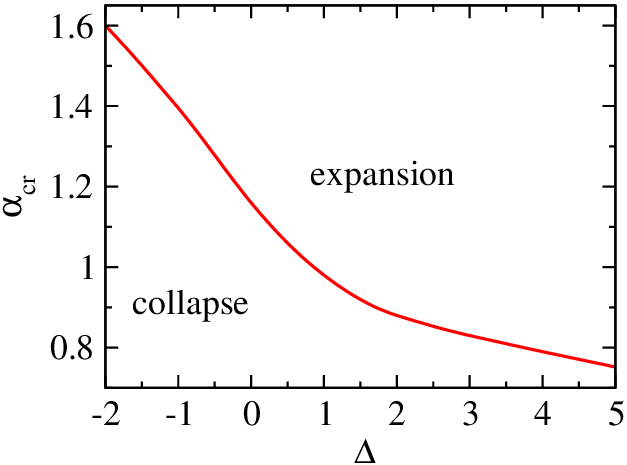}
\caption{(Color online) The stability diagram in the $(\alpha _{%
\mathrm{cr}},\Delta )$ plane for solutions stemming from the
single-component input based on Eq. (\ref{ux0vx0}). Other parameters
are $N=3$, $g_{1}=1,$ $g_{2}=0$.}
\label{fig:boundary}
\end{figure}

\begin{figure}[t]
\centering
\includegraphics*[width=0.45\textwidth]{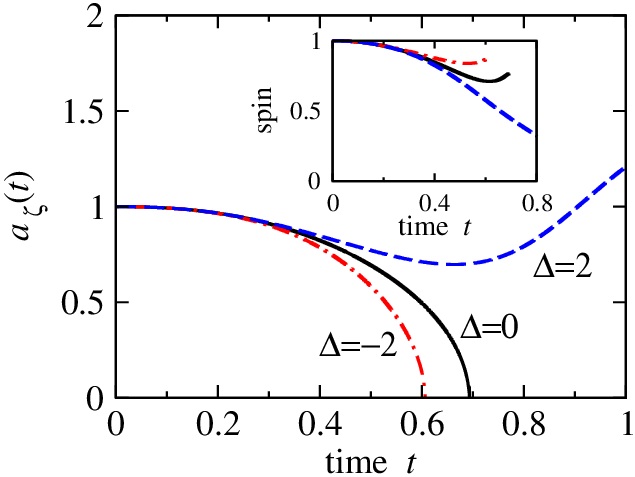}
\caption{(Color online) The time dependence of the width and mean value of the spin component $\langle\sigma_{z}\rangle$
of the collapsing and expanding wave packets, generated by input (\ref%
{ux0vx0}), at different values of the ZS strength, $\Delta $ (shown near the
curves). Other parameters are $N=3$, $g_{1}=1,$ $g_{2}=0$. }
\label{fig:spin}
\end{figure}

Regarding the effect of ZS on the collapse, there may be either a
competition or mutual enhancement of the SOC and ZS terms. The dependence of
the critical value, $\alpha _{\mathrm{cr}}$, on the ZS strength, $\Delta $,
is plotted in Fig. \ref{fig:boundary}, indicating a strong $\Delta
\leftrightarrow -\Delta $ asymmetry, related to the $\Delta $-dependence of
the Zeeman energy in Eq. \eqref{Zeeman}. In particular, the total energy
decreases at $\Delta <0$, which facilitates the onset of the collapse, hence
$\alpha _{\mathrm{cr}}$ is larger in this case. On the other hand, at $%
\Delta >0$ both the SOC\ and ZS terms resist the collapse, as seen in Fig. %
\ref{fig:spin}.

Another characteristic feature of the collapse-expansion dichotomy is the
time dependence of the spin presented in the inset of Fig. \ref{fig:spin},
which shows relatively slow non-decaying evolution if the collapse occurs,
and decays to zero otherwise (if the condensate expands, instead of blowing
up). The spin separation caused by the anomalous velocity leads to decrease
of the total spin, by producing a mixed spin state with $\sum_{j=x,y,z}%
\langle \sigma _{j}\rangle ^{2}<1.$ In particular, the state approximately
given by Eq. \eqref{eq:splitting} shows that, in the limit of $\alpha t\gg
a(0),$ all expectation components of the spin vanish, corresponding to the
maximally mixed spin state, where $\sum_{j=x,y,z}\langle \sigma _{j}\rangle
^{2}=0.$

\begin{figure}[t]
\centering
\includegraphics*[width=0.45\textwidth]{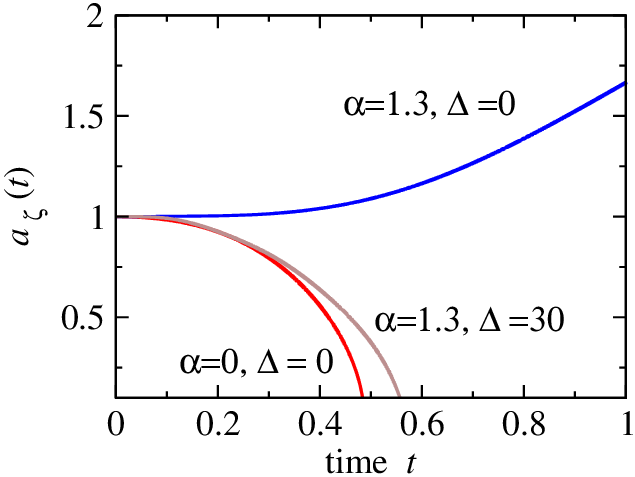}
\caption{(Color online) The time dependence of the width of the collapsing
or expanding wave packets, defined as per Eq. (\ref{zeta}), at
values of parameters shown near the plots. The simulations were performed
with initial conditions given by Eq. (\ref{ux0vx0}). At sufficiently
large $\Delta $ the collapse is enforced, the evolution of the width being
close to that in the absence of SOC. Note that $\alpha _{\mathrm{cr}%
}=1.16$ at $\Delta =0$. The norm fixed here is $N=3$. }
\label{fig:strong_D}
\end{figure}

At a sufficiently large $\Delta >0$, ZS can dominate in the spin evolution, making it
difficult to spatially separate the spin components.
This is shown in Fig. \ref{fig:strong_D}, where it is seen that,
in the limit of very large $\Delta$, the collapse occurs as in the absence
of SOC. The suppression of the SOC effect by large $\Delta $ can be
explained as follows \cite{Sherman2014}. Consider a Fourier component of the BEC
wave function with momentum $p.$ The corresponding spin precession rate, as
follows from Eq. \eqref{eq:Hspin}, is $\sqrt{\Delta ^{2}+4\alpha ^{2}p^{2}}$%
, and the orientation of the spin-precession axis is $\mathbf{n}_{s}=\left(
\alpha p,0,\Delta /2\right).$ In the limit of $\Delta \gg \alpha p,$ equivalent to
the condition $\Delta \gg \alpha/a(0),$ the
direction of $\mathbf{n}_{s}$ is close to the $z$-axis and the maximum
value of the anomalous velocity $\alpha\sigma_{x},$ achieved in the course of the BEC
evolution, corresponds to the maximally achieved $\sigma_{x}\sim 4\left(\alpha/a(0)\right)/\Delta,$
and becomes $\sim 4\alpha\left(\alpha/a(0)\right)/\Delta,$
decreasing as $1/\Delta.$ Therefore, in this limit, imposed on the value of $\Delta,$ the additional
condition, dependent on the BEC self-interaction, for the occurrence of the collapse can be rewritten as $4\alpha
\left(\alpha/a(0)\Delta\right)\ll V_{c},$ or, equivalently, $\Delta\gg 4\alpha^{2}/(a(0)V_{c}),$
where $V_{c}=\sqrt{\Lambda}/a(0)$ is given by Eq. \eqref{Vc}.

\subsection{The collapse driven by the cross-spin attraction: $%
g_{1}=0,g_{2}\neq 0,\Delta =0$}

To complete the analysis of the effects of SOC on the BEC collapse,
we consider the system of GPEs (\ref{eq:eq1}) and (\ref{eq:eq2}),
which includes solely the cross-nonlinearity, \textit{viz}., $g_{1}=0$ and $%
g_{2}=1$ at the initial condition $\langle\sigma_{z}\rangle=1,$ as
presented in Eq. \eqref{ux0vx0}. Here the dynamics is qualitatively different from the
evolution for the same-spin interaction since in the absence of the spin-orbit coupling
the initial state spreads for any norm. Therefore, the
collapse can happen only as a result of the SOC-driven spatial
splitting of the wave packet into spin-polarized complexes, similar to that
predicted by Eq. (\ref{eq:splitting}) resulting in the cross-spin attraction
between the spin components. Thus, in contrast to the above
case with $g_{2}=0$, here the collapse becomes possible when the SOC
strength \textit{exceeds} a certain critical value: $\alpha >\alpha _{%
\mathrm{cr}}$. It is worth noting that at $g_{1}=0$ in a state with $\langle\sigma_{x}\rangle=1,$
defined in Eq. \eqref{ux0vx01}, the critical norm $\widetilde{N}_{\rm cr}$ is
given by $\widetilde{N}_{\rm cr}=(3\pi)^{1/4}g_{2}^{-1/2},$
as can be seen from Eqs. \eqref{eq:Ekin},\eqref{crit}, and \eqref{eq:g1g2}.

\begin{figure}[t]
\centering
\includegraphics*[width=0.45\textwidth]{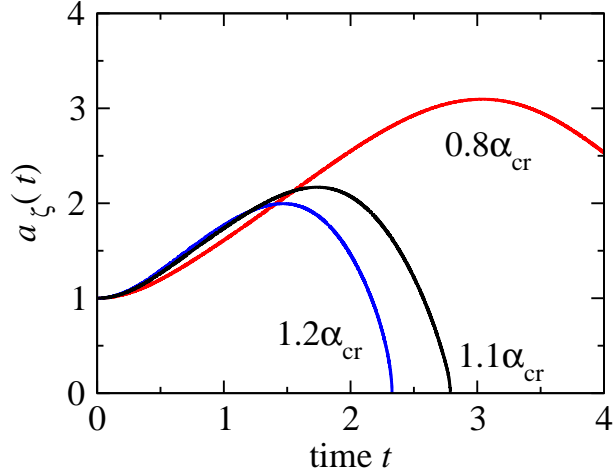}
\caption{(Color online) The widths of the collapsing wave packets,
calculated as per Eq. (\ref{zeta}), in the system which includes
solely the cross-attraction between the components, i.e., $g_{1}=0$ and $%
g_{2}=1$ in Eqs. (\ref{eq:eq1}) and (\ref{eq:eq2}), other
parameters being $N=6$ and $\Delta =0$, which corresponds to $%
\alpha _{\mathrm{cr}}=0.76$. The simulations were initiated by input (%
\ref{ux0vx0}).}
\label{fig:crosscoupling}
\end{figure}

Although the resulting evolution is very complex, the critical value can be estimated here from the following scaling
argument based on presentation of the wavefunction similar to that in Eq. \eqref{eq:splitting}.
First, let us assume that the spin-orbit coupling is sufficiently strong to ensure that
the initial wavepacket with the norm $N$ and $\langle\sigma_{z}\rangle=1$ rapidly splits in two
Gaussian wavepackets, having the same density profiles as the initial one, albeit with the norm $N/2$
and spin projections $\langle\sigma_{x}\rangle=\pm1.$ Then, if $N>2\widetilde{N}_{\rm cr},$ each spatial
component of the BEC will collapse independently. This argument, consistent with Eq. \eqref{eq:DeltaEg1g2},
shows that the minimal time required to switch on the cross-spin interaction is the time of the
essential spatial separation of the spin components, that is $\sim a(0)/\alpha$ \cite{Sherman2014}.
On the other hand, in the absence of the self-interaction, at time $t>a^{2},$ the wavepacket spreads
with the rate of the order of $1/a(0),$ due to the Heisenberg momentum-position
uncertainty ratio. Therefore, the required separation can be achieved at $\alpha\sim 1/a(0),$
establishing the low boundary for the critical $\alpha$ for the norm of the order of $\widetilde{N}_{\rm cr}.$
Note that for the norm $N\gg\widetilde{N}_{\rm cr},$ a strong separation is not needed for the collapse,
leading to a decrease in the $\alpha_{\rm cr}$ with $N.$ This realization requires a separate analysis.
The results of numerical calculations shown in Fig. \ref{fig:crosscoupling} confirm that this realization
self-interaction is indeed  qualitatively different from the one with the same-spin interaction. The numerical value
$\alpha _{\mathrm{cr}}=0.76$ agrees well with the above presented scaling argument $\alpha_{\mathrm{cr}}\sim 1.$

For experimental manifestations of the effects considered here, one might follow proposal of Ref. \cite{Altin2011}
by  adding the realizations of SOC and ZS terms. Since it is possible to tune the atomic
scattering length to change the interatomic interaction by Feshbach resonance \cite{Cornish2000},
one might reduce the strength of the repulsive two-body interaction and amplify
the three-body attractive interaction, making the three-body collisions dominant.
In addition, the three-body losses could be ignored in the system since the collapse
time scale is short.

\section{Conclusions}
\label{end}

We have studied the dynamics of density and spin patterns in the
one-dimensional BEC collapse driven by the quintic same- and
cross-spin interaction in the presence of artificial spin-orbit coupling (SOC) and Zeeman splitting
(ZS). The dynamics is determined by the interplay of the
nonlinear attraction and the splitting of the spinor wave function into
spin-polarized complexes, driven by the SOC-induced anomalous spin-dependent velocity [see
Eq. (\ref{eq:splitting})]. We performed
investigation of the qualitative features of the collapse dynamics, varying the SOC and ZS strengths. In
particular, we found that the Zeeman term can either support or suppress the collapse,
depending on the initial conditions and direction (sign) of the Zeeman
field. The effect of the SOC critically depends on the form of the quintic attraction
terms. Namely, it tends to suppress the collapse which is driven by the
self-attraction in each BEC component, and, on the other hand, SOC supports the
collapse in systems with the cross-spin attraction. We numerically obtained
critical values of the spin-orbit coupling, establishing the boundary between collapsing and
expanding states. In addition, our calculations demonstrated that due to a non-Gaussian evolution
of the density, to see the collapse in these systems,
one needs to study the inverse participation ratio rather than expectation values of the coordinate.
Taking into account continuous Lie symmetries \cite{VVLieS} can be useful for the future studies of the spin effects. 
Although topological arguments
related to the formation of vortex-like
structures, which appear due to the spin-orbit coupling \cite{Sakaguchi2014,Sakaguchi2016}
or without it \cite{WangPRA},
cannot be directly applied here, studies of possible stationary
pseudo-spinor structures in two dimensions, including relaxation effects,
are of interest and will be a topic of further research.

\section{Acknowledgement}

This work was partially supported by NSFC (11474193), SMSTC (18010500400 and
18ZR1415500), and the Program for Eastern Scholar in China. J.L. and W.L.
acknowledge support by Okinawa Institute of Science and Technology Graduate
University. E.Y.S. acknowledges support by the Spanish Ministry of Science and the European Regional
Development Fund through PGC2018-101355-B-I00 (MCIU/AEI/FEDER,UE), and the Basque Government through
Grant No. IT986-16. The work of B.A.M. is supported, in part, by Israel Science Foundation through grant No. 1287/17. X.C. thanks the Ram\'{o}n y Cajal grant (RYC-2017-22482).
We are grateful to Th. Busch for discussion and valuable comments.


\begin{thebibliography}{99}

\bibitem{Sulem1999} C. Sulem and P. L. Sulem, \textit{The nonlinear Schr\"{o}%
dinger Equation: Self-Focusing and Wave Collapse} (Springer: Berlin, 1999).

\bibitem{Fibich} G. Fibich, \textit{The Nonlinear Schr\"{o}dinger Equation:
Singular Solutions and Optical Collapse} (Springer: Heidelberg, 2015).

\bibitem{Carr2009} R. V. Mishmash and L. D. Carr, Phys. Rev. Lett. \textbf{103}, 140403 (2009).

\bibitem{Cornish2000} S. L. Cornish, N. R. Claussen, J. L. Roberts, E. A.
Cornell, and C. E. Wieman, Phys. Rev. Lett. \textbf{85}, 1795 (2000).

\bibitem{spielman2009} {Y.-J. Lin, R. L. Compton, K. Jim\'{e}nez-Garcia, J.
V. Porto and I. B. Spielman}, {Nature} \textbf{462}, {628} {(2009)}.

\bibitem{Dalibard2011} J. Dalibard, F. Gerbier, G. Juzeliunas, and P. \"{O}%
hberg, Rev. Mod. Phys. \textbf{83}, 1523 (2011).

\bibitem{wang2010} {C. Wang, C. Gao, C.-M. Jian, and H. Zhai}, {Phys. Rev.
Lett.} \textbf{105}, {160403} {(2010)}.

\bibitem{spielman2011} {Y.-J. Lin, K. Jim\'{e}nez-Garc\'{\i}a, and I. B.
Spielman}, {Nature} \textbf{471}, {83} {(2011)}.

\bibitem{Zhai2012} {H. Zhai}, {Int. J. Mod. Phys. B} \textbf{26}, {1230001} {%
(2012)}.

\bibitem{Spielman2013} {V. Galitski and I. B. Spielman}, {Nature} \textbf{494%
}, {49} {(2013)}.

\bibitem{Zhang2016} Y. Zhang, M. E. Mossman, Th. Busch,
P. Engels, and C. Zhang, Front. Phys. \textbf{11} 118103 (2016).

\bibitem{Konotop2005} V. V. Konotop and P. Pacciani, Phys. Rev. Lett.
\textbf{94}, 240405 (2005).

\bibitem{Dias2016} J.-P. Dias, M. Figueira, and V. V. Konotop, Stud. in
Appl. Math. \textbf{136} 241 (2016).

\bibitem{WangKdv} D.-S Wang and J. Liu, Appl. Math. Lett. \textbf{79}, 211 (2018).

\bibitem{Mardonov2015} Sh. Mardonov, E. Ya. Sherman, J. G. Muga, H.-W.
Wang, Y. Ban, and X. Chen Phys. Rev. A \textbf{91}, 043604 (2015).

\bibitem{Yu2017} Z.-F. Yu, A.-X. Zhang, R.-A. Tang, H.-P. Xu, J.-M.
Gao, and J.-K. Xue, Phys. Rev. A \textbf{95} 033607 (2017).

\bibitem{Sakaguchi2014} H. Sakaguchi, B. Li, and B. A. Malomed, Phys. Rev. E
\textbf{89}, 032920 (2014).

\bibitem{Sakaguchi2016} H. Sakaguchi, E.Ya. Sherman, and B. A. Malomed,
Phys. Rev. E \textbf{94}, 032202 (2016).

\bibitem{EPL} B. A. Malomed, EPL \textbf{122}, 36001 (2018).

\bibitem{Konler2002} T. K\"ohler, Phys. Rev. Lett. \textbf{89}, 210404
(2002).

\bibitem{Xi2016} K.-T. Xi and H. Saito, Phys. Rev. A \textbf{93}, 011604(R)
(2016).

\bibitem{Astrakharchik2005} G. E. Astrakharchik, J. Boronat, J. Casulleras,
and S. Giorgini, Phys. Rev. Lett. \textbf{95}, 190407 (2005).

\bibitem{Chiquillo2017} E. Chiquillo, J. Phys. A: Math. Theor. \textbf{50},
105001 (2017).

\bibitem{Maim} B. A. Malomed, A. I. Maimistov, and A. Desyatnikov, Phys.
Lett. A \textbf{254}, 179 (1999).

\bibitem{Abdullaev2005} F. Kh. Abdullaev and M. Salerno, Phys. Rev. A
\textbf{72}, 033617 (2005).

\bibitem{Anderson} D. Anderson, Phys. Rev. A \textbf{27}, 3135 (1983).

\bibitem{Progress} B. A. Malomed, Progr. Optics \textbf{43}, 71 (2002).

\bibitem{Ermakov} V. P. Ermakov, Universitetskie Izvestiya, Kiev, No. 9, 1 (1880) (in Russian).

\bibitem{Ermakov2} W. K. Schief, C. Rogers, and A. P. Bassom, J. Phys. A:
Math. Gen \textbf{29}, 903 (1996).


\bibitem{Abdullaev2003} F. Kh. Abdullaev, J. G. Caputo, R. A. Kraenkel, and
B. A. Malomed, Phys. Rev. A \textbf{67}, 013605 (2003).

\bibitem{Manakov1974} S. V. Manakov, {Sov. Phys. JETP} \textbf{38}, 248
(1974).

\bibitem{Chaves2015} A. Chaves, G. A. Farias, F. M. Peeters, and R. Ferreira,
Commun. in Comp. Phys. \textbf{17}, 850 (2015).

\bibitem{IPR2000} F. Evers and A. D. Mirlin, Phys. Rev. Lett. \textbf{84},
3690 (2000).

\bibitem{Sherman2014} E. Ya. Sherman and D. Sokolovski, New Journ. of
Phys. \textbf{16}, 015013 (2014).

\bibitem{Altin2011} P. A. Altin, G. R. Dennis, G. D. McDonald, D. D\"{o}ring, J. E. Debs, J. D. Close, C. M. Savage, and N. P. Robins, Phys. Rev. A \textbf{84}, 033632 (2011).
%
\bibitem{VVLieS} J. Belmonte-Beitia, V\'{i}ctor M. P\'{e}rez-Garc\'{i}a, and V. Vekslerchik, Phys. Rev. Lett. \textbf{98}, 064102 (2007).

\bibitem{WangPRA} D.-S Wang, S.-W Song, B. Xiong, and W. M. Liu, Phys. Rev. A \textbf{84}, 053607 (2011).










\end{thebibliography}
\end{document}